\documentclass[prl,aps,twocolumn,showpacs]{revtex4}
\usepackage{epsfig}

\draft

\begin{document}
\author{Li-Bin Fu,$^{1,2}$ Shi-Gang Chen,$^{1}$ and Bambi Hu$^{3}$}
\address{$^1$Institute of Applied Physics and Computational Mathematics,\\
P.O. Box 8009 (28), Beijing 100088, P.R. China}
\address{$^2$Max-Planck-Institute for the Physics of Complex systems,\\
N\"{o}thnitzer Str. 38, 01187 Dresden, Germany}
\address{$^3$Department of Physics and Centre for Nonlinear Studies, Hong\\
Kong Baptist University , Hong Kong, P.R. China}
\title{Landau-Zener tunneling with many-body quantum effects in crystals of molecular magnets}
\date{\today }

\begin{abstract}
We present a quantum interpretation of the heights in hysteresis
of $Fe_{8}$ molecule at lower temperatures by treating the crystal
as an Ising spin system with the dipolar interaction between
spins. Then we apply it to two limit cases : rapid and adiabatic
regions. Our theoretical analysis is in agreement with the
experimental observation in these regions, which indicates that
the steps in hysteresis loops of magnetization of Fe$_{8}$ at
lower temperatures show a pure quantum process.
\end{abstract}

 \maketitle

\section{Introduction}

Crystals of molecular magnets, such as Fe$_{8}$ and Mn$_{12}$, \ have
attracted much attention for their connection to many macroscopic quantum
phenomena \cite{mn12,ljk,all,lzsci}. They may also have important
applications in magnetic memory and quantum computing \cite{lzsci,qc}. The
earliest and most spectacular observation on such a system is the quantum
steps in the hysteresis loops of magnetization at low \ temperatures \cite
{mn12,lzsci}. These quantum steps are the manifestations of macroscopic
quantum tunneling, resulting from the tunneling between different spin
states of large molecular spins ($S=10$ for both Fe$_{8}$ and Mn$_{12}$, $%
S=9/2$ for Mn$_{4}$). Fe$_{8}$ is particular interesting because
the steps in the hysteresis will become temperature independent
below $0.36K,$ which shows a pure tunneling process \cite{lzsci}.
This tunneling phenomenon is complicated by the interaction
between spins and other environmental effects. Intensive efforts
have been devoted to explain the step features via different
approaches \cite{liu8,new1,new2}. The modifications of other
environmental effects have also been studied, such as the nuclear
spin effects \cite{ham,new3,new4}.

More recently, Liu, et al presented a successful theory on the
height of quantum steps in the hysteresis loop when the
temperature is low enough that the thermal effects can be
neglected \cite{liu}. By treating Fe$_{8}$ crystal as a system of
Ising spins sitting at each site of the lattice, the step heights
measured in experiment were successful reproduced by directly
solving an evolution equation described by the dipolar fields
distribution. The results have been compared to Landau-Zener (LZ)
model \cite{lz}, which has been used to extract the tunnel
splitting $\Delta $ of a single molecular spin from step heights
\cite{lzsci,liu10}. In their simulation, the dipolar interaction
between spins is treated by a mean-field theory and the flipping
of each spin is independently.

In this paper, we also model the Fe$_{8}$ crystal as a system of
Ising spins sitting at each site of the lattice and taking into
account the dipolar interaction between spins. But different form
Ref. \cite{liu}, we treat the system as a quantum many-body system
and formally give a formula to evaluate magnetization. Although it
can not be used to calculate magnetization in most cases for the
algorithm reason (the time of calculating is increasing
exponentially with the number of sites), we apply it to two limit
cases: rapid and adiabatic cases. Through the pure quantum
approach, our theory successfully interpret the quantum step
heights in these two regions. As an application of our theory, we
show that the tunnel splitting $\Delta _{e}$ measured with the LZ
model \cite{lzsci,liu10} is proportional to the true tunnel
splitting $\Delta $ of a single molecular with a geometry factor
(depending on shape and lattice structure). This result has also
been obtained in Ref. \cite{liu}. In the adiabatic limit, we show
the measured tunnel splitting $\Delta _{e}$ is dependent with the
sweeping rates with the power law: $\Delta _{e}\sim \alpha
^{1/2}$. This prediction agrees with the experimental
observations.

\section{Model}

We model the Fe$_{8}$ crystal as spin lattices with the realistic constants $%
10.52:$ $14.05:$ $15.00$ $(A)$ and angles $89.9^{\circ }:$
$109.6^{\circ }:109.3^{\circ }\;$ between the axes, which is a
triclinic lattice; the shortest axis $a$ as the easy axis
(actually there is an angle of about $8^{\circ }$ between them,
but it does not affect significantly the results). As in Ref.
\cite{liu}, we focus on one
step for simplicity, that is, the tunneling between the two lowest levels $%
(S_{z}=\pm 10)$. The effective Hamiltonian operating in the subspace is \cite
{ham}
\begin{equation}
{\cal H}=-\sum\limits_{i}g\mu _{B}S\mu _{0}H\sigma _{z}^{(i)}-\frac{1}{2}%
\sum\limits_{ij}V_{ij}\sigma _{z}^{(i)}\sigma _{z}^{(j)}+\frac{1}{2}%
\sum_{i}\Delta \sigma _{x}^{(i)}.  \label{ham}
\end{equation}
The first term describes the Zeeman energy, and $H$ is the external field
applied in the direction of the easy axis. The second term is the spins
interaction with dipolar potential $V_{ij}=E_{d}(3\cos ^{2}\theta -1)\Omega
_{0}/r_{ij}^{3},$ $E_{d}=\frac{\mu _{0}}{4\pi }(g\mu _{B}S)^{2}/\Omega _{0}$%
, where $\vec{r}_{ij}$ is the displacement between the spins, $\theta $ is
the angle between $\vec{r}_{ij}$ and the easy axis, and $\Omega _{0}$ is the
unit cell volume. The last term describes tunneling and $\Delta $ is the
tunnel splitting. $\sigma _{z}$ and $\sigma _{x}$ are Pauli matrices, and $%
\{i\},\{j\}$ label molecular sites.

Because $\Delta /E_{d}\sim 10^{-6}$ (see Ref. \cite{exp}), the last term of
Eq. (\ref{ham}) can be regarded as a small perturbation, denoted as $W=\frac{%
1}{2}\sum_{i}\Delta \sigma _{x}^{(i)}.$ \ We can know the eigenstates of the
unperturbed system are $\phi _{\mu }=|\left. s_{1}^{\mu },s_{2}^{\mu
},\cdots ,s_{N}^{\mu }\right\rangle $ $(\mu =1,\cdots ,2^{N}),$ in which $%
s_{i}^{\mu }=\pm 1$ corresponding to the spin on site $i$ up and down
respectively. The eigenvalue for $\phi _{\mu }$ is
\begin{equation}
E_{\mu }=-g\mu _{B}S\mu _{0}H\sum\limits_{i}s_{i}^{\mu }-\frac{1}{2}%
\sum\limits_{ij}V_{ij}s_{i}^{\mu }s_{j}^{\mu }.
\end{equation}
It is easy to see that the energy levels would be degenerate when some sites
are geometric equivalent.

On the other hand, because $\sigma _{x}^{(i)}$ operating on $\phi _{\mu }$
will make the spin on the $i$-th site flip, the perturbation term, i.e., the
off-diagonal element, ${\cal H}_{\mu \nu }=\phi _{\mu }{\cal H}\phi _{\nu
}=\phi _{\mu }W\phi _{\nu }$, is not zero if and only if $s_{i}^{\mu
}=s_{i}^{\nu }$ $(i=1,2,...,j-1;j+1,j+2,...,N)$ and $s_{j}^{\mu }\neq
s_{j}^{\nu },$ where $j$ could be any site. At this time ${\cal H}_{\mu \nu
}=\Delta /2$.

From the perturbation theory \cite{qm}, we can know, if off-diagonal element
of two states is nonzero, the corresponding energy levels must have an
avoided crossing with a gap proportional to the off-diagonal element $\Delta
/2$. The gap of the avoided crossing is determined by the degenerate
properties of the levels. For example, if both levels are non-degenerate,
the gap is $\Delta ;$ but if one of them is two-fold degenerate, the gap is $%
\sqrt{2}\Delta ,$ and so on$.$ Higher order perturbations are much small, so
they can be regarded as crossings. For example, for the second order, the
gap is about $\Delta ^{2}/E_{d}\sim 10^{-6}\Delta ,$ so it can be treated as
a crossing.

\section{Magnetization and tunnel splitting }

\subsection{Result of our model}

Supposing the crystal has $N$ Fe$_{8}$ molecules, it is initially on the
state $|\left. -1,-1,\cdots ,-1\right\rangle $ in a large negative field,
then sweep the field with a constant rate $\alpha $ to the positive. Over an
avoided crossing, the spin involved will flip with the probability $%
1-p_{\delta }$ where $p_{\delta }=e^{-\frac{\pi (\delta \Delta )^{2}}{%
2\alpha }}$ if the gap is $\delta \Delta .$ The magnetization can be
formally expressed as
\begin{eqnarray}
M &=&-NSP_{-N}-(N-2)SP_{-(N-1)}+  \nonumber \\
&&+\cdots -(N-2i)SP_{-(N-i)}+\cdots ,  \label{m}
\end{eqnarray}
where $P_{-(N-i)}$ is the sum of probability for all the energy levels with $%
i$ spins up. For example, assuming there are $m$ levels of one-spin up
states, which are denoted as $E_{-(N-1)}^{l}$ $(l=1,2,\cdots ,m).$ The
energy structure is shown in Fig.1 where only the initial level $E_{-N}$ and
one-spin up levels are plotted. The energy gap between $E_{-(N-1)}^{l}$ and $%
E_{-N}$ is $\delta _{1}^{l}\Delta .$ Then we can obtain the probability of
the initial level,
\begin{equation}
P_{-N}=\prod_{l}^{m}p_{\delta _{1}^{l}}=e^{-\frac{\pi \Delta ^{2}}{2\alpha }%
c},  \label{pn}
\end{equation}
where $c=\sum_{i=1}^{m}(\delta _{1}^{i})^{2}.$ In analog, we can evaluate
the probability for any level involved in principle. It can be formally
expressed as
\begin{equation}
P_{-(N-l)}\sim \sum_{\alpha }\left[ \prod_{\beta =1}^{l}(1-p_{\delta _{\beta
}})\prod_{i}^{m_{\alpha }}p_{\delta ^{i}}\right] ,  \label{pni}
\end{equation}
in which we have ignored some subscripts for convenience.


\begin{figure}[!htb]
\begin{center}
\rotatebox{-90}{\resizebox *{6.0cm}{8.0cm} {\includegraphics
{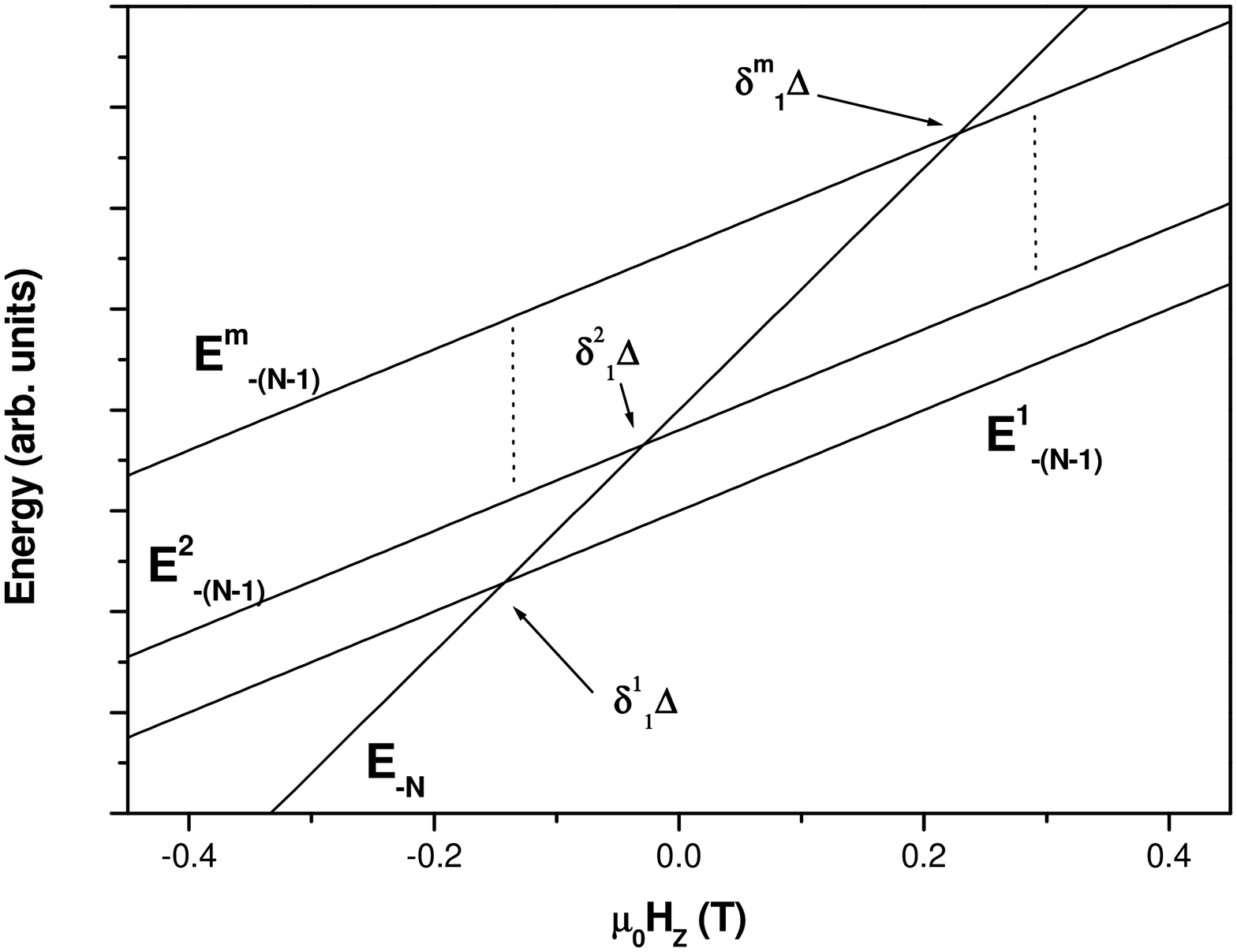}}} \caption{The energy structure of the initial level
$E_{-N}$ and one-spin up levels. The energy gap between
$E_{-(N-1)}^{l}$ and $E_{-N}$ is $\delta _{1}^{l}\Delta .$}
\end{center}
\label{fig1}
\end{figure}

There should be a term in Eq. (\ref{m}) corresponding to the adiabatic path:
starting from the initial level and keeping on the continuous branch at
every avoided crossing encountered. The probability of the adiabatic path
can be expressed as
\begin{equation}
P_{ad}=(1-p_{\delta _{1}^{1}})(1-p_{\delta _{2}^{1}})\cdots (1-p_{\delta
_{K}^{1}}),  \label{ad}
\end{equation}
where $(1-p_{\delta _{i}^{1}})$ is the probability of spin flipping over
each avoided crossing and $K$ is just the total number of the flipped spins
along this path.

Indeed, (\ref{m}) can not be calculated for nowaday computer when
the number of spins is large. But we can evaluate it for limit
cases.

For the high sweeping rate limit, $P=e^{-\frac{\pi \Delta ^{2}}{2\alpha }%
}\rightarrow 1,$ so $(1-P)$ is an infinitesimal. Form Eq. (\ref{pni}), one
knows that $P_{-(N-l)}$ is as the same order as $(1-P)^{l}$. Hence, to the
first order of $(1-P)$, the magnetization can be approximately expressed as
\begin{equation}
M\approx -NSP_{-N}-(N-2)SP_{-(N-1)}.
\end{equation}
Of course, the total probability must be conserved, i.e.,
\begin{equation}
P_{-N}+P_{-(N-1)}+\cdots +P_{-(N-i)}+\cdots =1.  \label{pp}
\end{equation}
So, to the first order of $(1-P)$, we have $P_{-N}+P_{-(N-1)}\approx 1,$
i.e., $P_{-(N-1)}\approx 1-P_{-N}.$ Substituting (\ref{pn}) into the above
formula, we get
\begin{equation}
M\approx -NS+2cS\frac{\pi \Delta ^{2}}{2\alpha }.  \label{mmm}
\end{equation}

In Refs. \cite{lzsci,liu10,exp}, Wernsdorfer and coworkers extract
the tunnel splitting $\Delta $ of a single molecular spin from the
magnetization by employing LZ model \cite{lz}. Based on LZ model,
the measured tunnel splitting $\Delta _{e}$ was calculated by the
following formula \cite{exp},
\begin{equation}
\Delta _{e}=\sqrt{-\frac{2\alpha }{\pi }\ln \left( \frac{1-M/M_{s}}{2}%
\right) },  \label{deltae}
\end{equation}
in which $M_{s}=NS.$

Substituting the theoretical predication of magnetization (\ref{mmm}) into
the above formula, we obtain
\begin{equation}
\Delta _{e}\simeq C\Delta ,  \label{ede}
\end{equation}
in which $C=\sqrt{\frac{c}{N}}.$

This result shows that for the rapid sweeping rate limit, the
measured tunnel splitting $\Delta _{e}$ is a constant, which
consists with the experiment observation \cite{exp}. Eq.
(\ref{ede}) also implies that $\Delta _{e}$ is not the true tunnel
splitting $\Delta $, but proportion to $\Delta $ with a factor $C$
which is dependent only on the geometry of the sample: its shape
and lattice structure. This consists with the result presented in
Ref. \cite{liu}.

If the sweeping rate is very small (adiabatic limit), $p_{\delta
_{i}^{j}}\rightarrow 0$, so $P_{-(N-i)}\rightarrow 0$ except for the
adiabatic term $P_{ad}\rightarrow 1,$ so $M\rightarrow M_{ad}=$ $-(N-2K)S$,
then from (\ref{deltae}) we obtain
\begin{equation}
\Delta _{e}=\sqrt{-\frac{2\alpha }{\pi }\ln \left( \frac{1-M_{ad}/M_{s}}{2}%
\right) }=k\alpha ^{1/2},  \label{dd}
\end{equation}
in which $k=\sqrt{\frac{2}{\pi }\ln \left( \frac{1-M_{ad}/M_{s}}{2}\right) }.
$ This shows that for the slow sweeping rates, the measured tunnel splitting
is strongly dependent on the sweeping rate. In the adiabatic limit it shows
a $1/2$ power law of the function of the sweeping rate: $\Delta _{e}=k\alpha
^{1/2}$. This feature is first revealed in this paper.

\subsection{Comparing our results with experiments}

For the high sweeping rate limit, as shown in Eq. (\ref{ede}), the
measured tunnel splitting is a constant, which consists with
experimental observation. For the adiabatic region, we can
calculate the measured tunnel splitting in the adiabatic limit
since the number of levels involved in adiabatic path is
proportion to the number of spins $N$. We can find the adiabatic
path by following the adiabatic process: starting from the initial
level and keeping the state on the continuous branch at every
avoided crossing encountered. In Fig. 2, we plot the adiabatic
magnetization $M_{ad}$ of $n\times n\times n$ lattice for
different total number of spins. One can find that as the total
number of spins is large enough, $M_{ad}$ becomes independent on
the total number, and tends to a constant $M_{ad}/M_{s}\simeq
-0.29$, i.e., $k=0.528$ . We also calculate $M_{ad}$ for the case
of $(a\times b\times c):16\times 8\times 8$ lattice ($a$ is easy
axis direction), and obtain $M_{ad}/M_{s}=-0.37$, i.e., $k=0.49$.
In Fig. 3, we compare the theoretical evaluation with the
experiment observation \cite{exp}. It is shown that for three different $%
Fe_{8}$ isotopes, the three curves of $\Delta _{e}$ tend to merge together
with the same tendency $\Delta _{e}\propto \alpha ^{1/2}$ . This tendency
consists with our theoretical prediction. We argue that the adiabatic
evolution of the system is only determined by the levels structure, so we
cannot read the information of the tunnel splitting from the adiabatic
process. This feature can be found in Fig. 5 where three curves of different
isotopes have the same tendency.


\begin{figure}[!htb]
\begin{center}
\rotatebox{-90}{\resizebox *{6.0cm}{8.0cm} {\includegraphics
{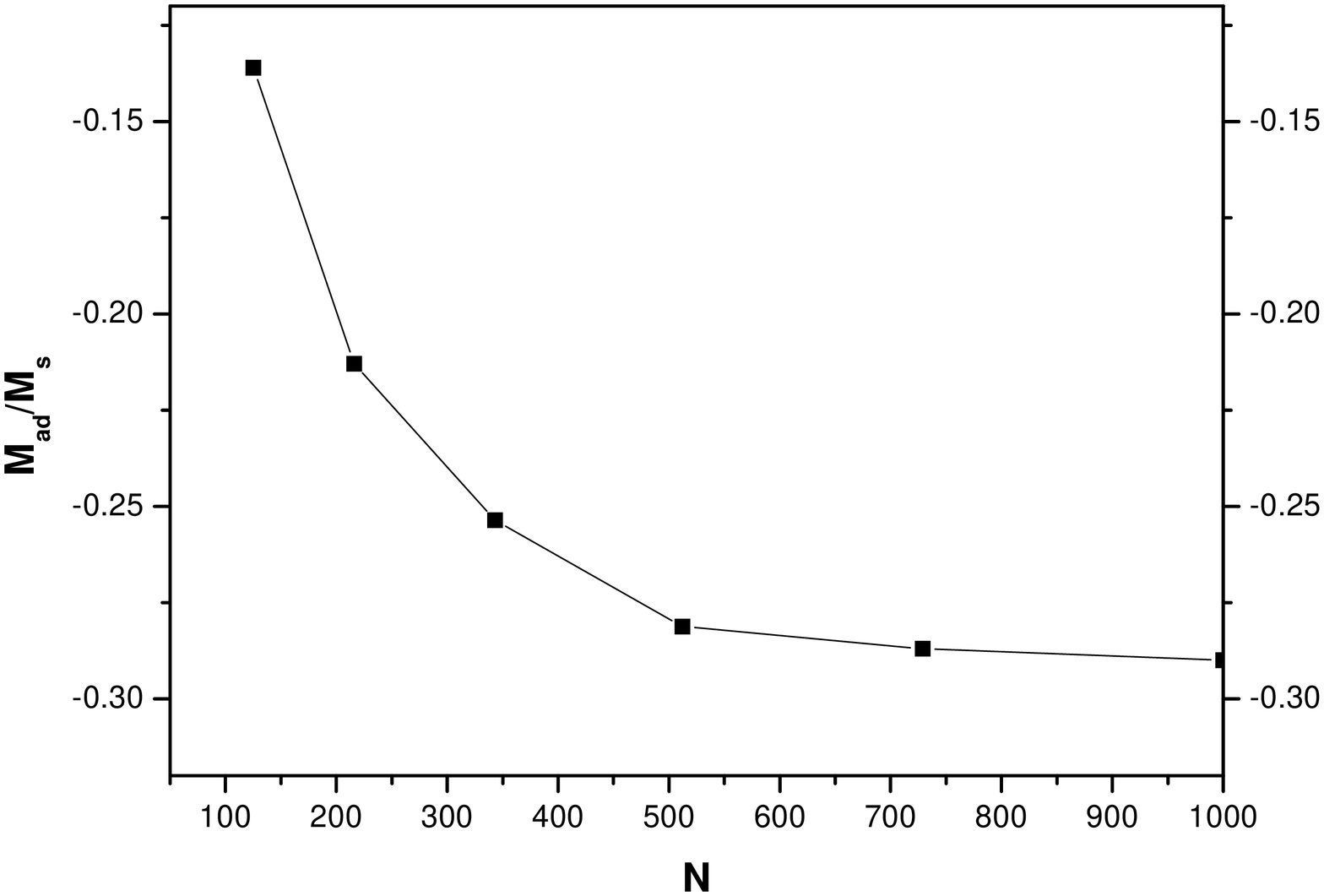}}} \caption{The adiabatic magnetization of $n\times
n\times n$ $Fe_{8}$ triclinic crystals for different number of
spins. The solid line is guide for eyes. $N=n\times n\times n$ is
the total number of spins.}
\end{center}
\label{fig2}
\end{figure}


\begin{figure}[!htb]
\begin{center}
\rotatebox{-90}{\resizebox *{6.0cm}{8.0cm} {\includegraphics
{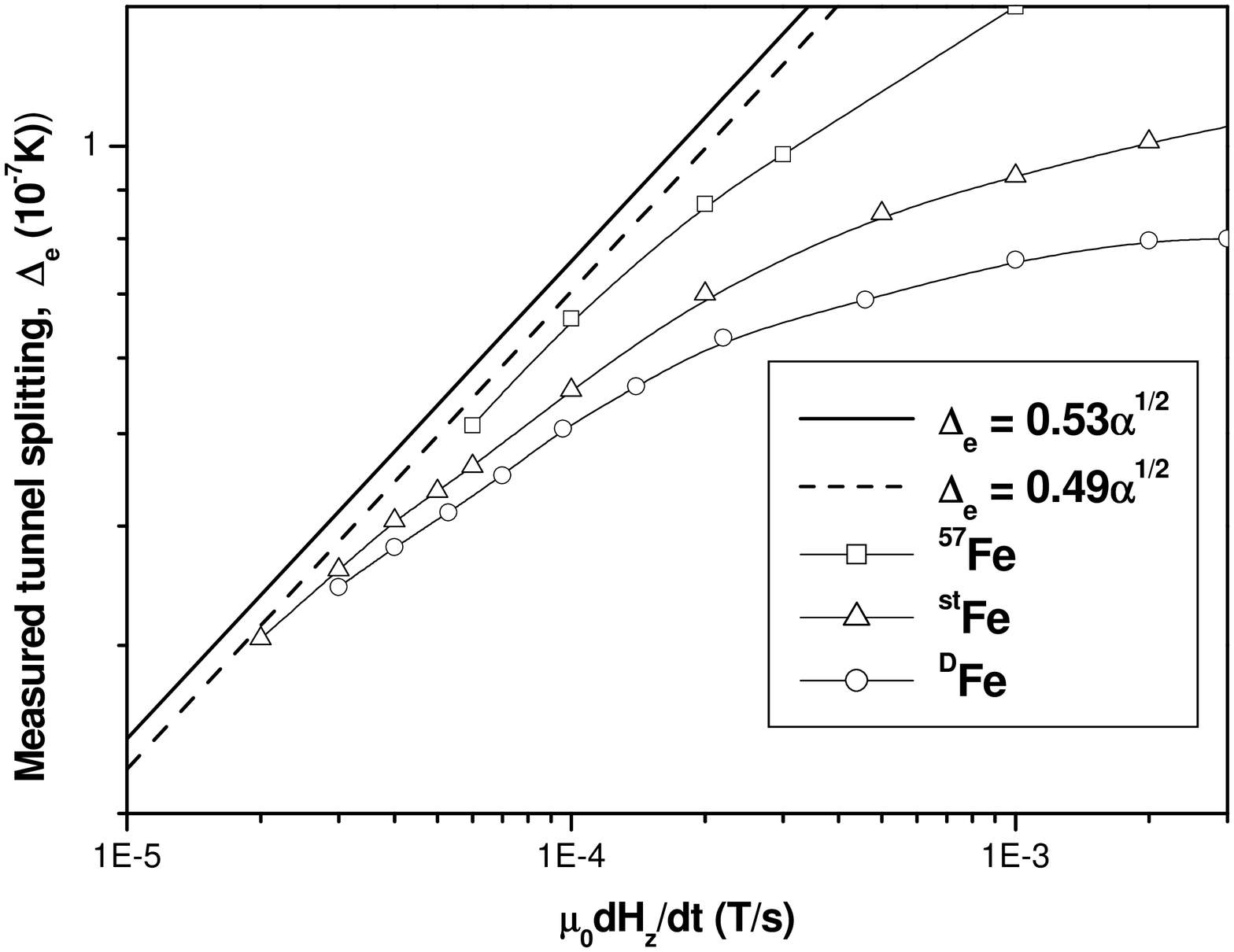}}} \caption{$\Delta _{e}$ of $Fe_{8}$ crystals for small
sweeping rates. The solid line is theory prediction with $n\times
n\times n$, the dashed line is for $2n\times n\times n$, the
others are the experimental data for different $Fe_{8}$ isotopes
[15].}
\end{center}
\label{fig3}
\end{figure}

\section{Conclusion}

In summary, we have given a pure quantum interpretation of the step heights
in hysteresis loops of $Fe_{8}$ molecule by treating the crystal as a system
of Ising spins sitting at each site of the lattice with the dipolar
interaction between spins. Our theoretical analysis is in agreement with the
experimental observation in the rapid and adiabatic limits. For the rapid
sweeping rates, we show that the measured tunnel splitting $\Delta _{e}$ is
a constant which is proportional to the tunnel splitting of the single
molecular spin, i.e., $\Delta _{e}=C\Delta $. The factor $C$ depends on the
sample geometry. But for the adiabatic limit, the magnetization becomes
independent on the sweeping rate, and tends to a constant. This feature
leads to that the measured tunnel splitting $\Delta _{e}$ is strongly
dependent on the sweeping rate $\alpha $, and be a $1/2$ power law of
sweeping rate: $\Delta _{e}\sim \alpha ^{1/2}$.

\section*{Acknowledgments}

This work was supported in part by the grants from the 973 Project
of China and the Hong Kong Research Grants Council (RGB). Dr. L.-B
Fu acknowledges the support of the Alexander von Humboldt
Foundation.

\end{document}